\documentclass[AMA,Times2COL]{WileyNJDv5} 

\articletype{Article Type}%

\received{Date Month Year}
\revised{Date Month Year}
\accepted{Date Month Year}
\journal{Journal}
\volume{00}
\copyyear{2023}
\startpage{1}

\usepackage{xr}

\usepackage{amssymb}
\usepackage{amsmath}
\usepackage{dsfont}
\usepackage{multirow}
\usepackage{latexsym}
\usepackage{epsfig}
\usepackage{graphicx}
\usepackage{epstopdf}
\usepackage{float}
\usepackage{flushend}

\usepackage{subcaption}
\usepackage{caption}
\usepackage{amsfonts}
\usepackage{xcolor}

\def\boxit#1{\vbox{\hrule\hbox{\vrule\kern6pt
          \vbox{\kern6pt#1\kern6pt}\kern6pt\vrule}\hrule}}

\usepackage{comment}

\newcommand{\bfa}{{\bf a}}
\newcommand{\bfb}{{\bf b}}

\newcommand{\bfm}{{\bf m}}

\newcommand{\bfp}{{\bf p}}

\newcommand{\bfv}{{\bf v}}

\newcommand{\bfx}{{\bf x}}
\newcommand{\bfy}{{\bf y}}
\newcommand{\bfz}{{\bf z}}

\newcommand{\bfA}{{\bf A}}
\newcommand{\bfB}{{\bf B}}
\newcommand{\bfC}{{\bf C}}
\newcommand{\bfD}{{\bf D}}

\newcommand{\bfL}{{\bf L}}
\newcommand{\bfM}{{\bf M}}
\newcommand{\bfN}{{\bf N}}

\newcommand{\bfQ}{{\bf Q}}

\newcommand{\bfS}{{\bf S}}
\newcommand{\bfT}{{\bf T}}

\newcommand{\bfV}{{\bf V}}

\newcommand{\bfX}{{\bf X}}
\newcommand{\bfY}{{\bf Y}}
\newcommand{\bfZ}{{\bf Z}}

\newcommand{\RNum}[1]{\uppercase\expandafter{\romannumeral #1\relax}}

\newcommand{\bftheta}{\mbox{\boldmath $\theta$}}

\newcommand{\bfmu}{\mbox{\boldmath $\mu$}}

\newcommand{\bfPsi}{\mbox{\boldmath $\Psi$}}
\newcommand{\bfsigma}{\mbox{\boldmath $\sigma$}}

\newcommand{\bftau}{\mbox{\boldmath $\tau$}}

\newcommand{\bfOmega}{\mbox{\boldmath $\Omega$}}

\newcommand{\diag}{\mathrm{diag}}

\newcommand{\bfzero}{{\bf 0}}

\begin{document}



%
%

\title{Bayesian Clustering Factor Models}


\author[1]{Hwasoo Shin, PhD}
\author[2]{Marco A. R. Ferreira, PhD}
\author[3]{Allison N. Tegge, PhD}

\address[1]{\orgname{Henry Ford Health}, \orgaddress{Detroit, MI, 48202, USA}}
\address[2]{Department of Statistics, \orgname{Virginia Tech},  \orgaddress{Blacksburg, VA, 24061, USA}}
\address[3]{Fralin Biomedical Research Institute, \orgname{Virginia Tech}, \orgaddress{Roanoke, Virginia, 24016, USA}}
\corres{Allison N. Tegge PhD, Fralin Biomedical Research Institute, Virginia Tech, Roanoke, Virginia, 24016, USA; ategge@vt.edu}






\abstract[Abstract]{We present a novel framework for concomitant dimension reduction and clustering. This framework is based on a novel class of Bayesian clustering factor models. These models assume a factor model structure where the vectors of common factors follow a mixture of Gaussian distributions.  We develop a Gibbs sampler to explore the posterior distribution and propose an information criterion to select the number of clusters and the number of factors. Simulation studies show that our inferential approach appropriately quantifies uncertainty. In addition, when compared to a previously published competitor method, our information criterion has favorable performance in terms of correct selection of number of clusters and number of factors. Finally, we illustrate the capabilities of our framework with an application to data on recovery from opioid use disorder where clustering of individuals may facilitate personalized health care. }

\keywords{Bayesian factors models, Clustering methods, Mixtures of Gaussian distributions}
%



\maketitle

\renewcommand\thefootnote{}

\renewcommand\thefootnote{\fnsymbol{footnote}}
\setcounter{footnote}{1}

\section{Introduction}\label{sec:Introduction}

We consider data from cross-sectional studies where subjects respond to a long questionnaire consisting of multiple, often related, assessments. In these studies, clustering of subjects based on their responses may facilitate the development and implementation of personalized health care. Unfortunately, medical use of results from clustering of multivariate data with a large number of dimensions is challenging because often the clusters are difficult to interpret and, thus, fail to provide actionable information. In the past, researchers have 
combined principal component analysis (PCA) to estimate latent structures and, based on these structures, $k$-means clustering \citep{hastie2009elements} to find clusters \citep{yeung2001principal, craft:shin:tegge:2022}. However, this combination of PCA and $k$-means, henceforth referred to as PCA+$k$, does not quantify uncertainty in the estimates of parameters and in the assignment of subjects to clusters. In addition,  PCA+$k$ relies on heuristics to choose the number of principal components and the number of clusters. To address these limitations, we propose novel Bayesian clustering factor models (BCFM) that assume clustering occurs in a latent space, thereby providing meaningful interpretation for each of the clusters. Our proposed framework combines factor models and mixtures of Gaussian distributions for concomitant dimension reduction and clustering.

We develop an inferential framework for parameter estimation, selection of number of factors and number of clusters, and cluster assignment of subjects. Our inferential framework uses conditionally conjugate priors that allow flexibility in the incorporation of prior information, and facilitate computation. In addition, we propose a default choice of prior hyperparameters inspired by unit information priors \citep{kass:1995,steele2010performance} that allow data analysts to automatically use our BCFM framework without expert knowledge of prior elicitation. We develop a Markov chain Monte Carlo (MCMC) algorithm \citep{gelfand:smith:1990, robe:case:2005,game:lope:2006}  to explore the posterior distribution of parameters. Specifically, the MCMC algorithm we propose is a Gibbs sampler based on the full conditional distributions of the unknown quantities of the model. Finally, we propose an information criterion that uses the MCMC output to ---within a model selection framework--- select the optimal number of factors and number of clusters.

To examine the accuracy of the BCFM estimation framework that we propose, we have performed simulation studies. Specifically, to evaluate the quality of the proposed estimation approach and the identifiability of the model, we simulated a dataset with three factors and four clusters. We then applied our MCMC-based estimation approach, which was able to provide parameter estimates close to the true values. In addition, the 95\% credible intervals included the true values of the parameters more than 95\% of the time. Further, about 96\% of the subjects were assigned to their correct clusters. Taken together, these results show appropriate uncertainty quantification provided by our proposed BCFM estimation and clustering approaches. 

To evaluate the quality of the proposed information criterion for the selection of number of factors and number of clusters in BCFM, we considered 10 different settings of separation among the clusters. For each separation setting, we simulated 100 datasets and for each dataset we obtained the best model ---the best number of factors and number of clusters---  based on PCA+$k$ and on our proposed information criterion. Then, for each separation setting and for each model selection method, we computed the mean number of factors and the mean number of clusters of the chosen model. According to these measures, when compared to PCA+$k$, model selection with our information criterion has favorable performance in terms of selection of number of clusters and number of factors much closer to the correct number of clusters and number of factors. 
Finally, we illustrate the capabilities of our BCFM framework with an application to data on recovery from opioid use disorder where clustering of individuals may facilitate personalized health care.

The remainder of the paper is organized as follows. Section \ref{Model_Specification} introduces the Bayesian clustering factor models that we propose. In Section \ref{sec:Statistical_Inference}, we propose an MCMC algorithm for the exploration of the posterior distribution and an information criterion for the selection of number of clusters and number of factors. Section~\ref{Simulation_Study} presents the results of simulation studies that evaluate the performance of our proposed inferential procedure in terms of estimation and model selection. Section~\ref{sec:OUD_Recovery_Data}  illustrates the capabilities of our BCFM framework with an application to data on recovery from opioid use disorder. 
Finally, Section \ref{sec:Conclusions} provides a brief discussion of the main contributions of this paper and possible avenues for future research.

\section{Model Specification}\label{Model_Specification}

\subsection{Bayesian Clustering Factor Models}\label{Bayesian_Clustering_Factor_Model}

To introduce BCFM, we consider a multivariate setting with $R$ variables observed for each of $n$ subjects. In addition, we assume that the dependence structure among the $R$ variables may be explained by a much smaller number $F$ of latent factors. Further, we assume that in this smaller $F$-dimensional latent space, the subjects may be clustered into $K$ clusters. This section describes the proposed BCFM for performing this concomitant dimension reduction and clustering.

Let $\bfy_{i}$ be an $R$-dimensional vector with the $R$ observed variables from subject $i$. BCFM assumes that the vector of observations follows the factor model
\begin{equation}\label{MainEQ1}
\bfy_{i} = \bfB \bfx_{i}  + \bfv_{i},
\end{equation}
where $\bfB$ is an $R \times F$ matrix of factor loadings, and $\bfx_{i}$ is an $F$-dimensional vector of common factors for subject $i$. The error vectors $\bfv_1, \ldots, \bfv_n$ are independent and identically distributed with $\bfv_{i} \sim \bfN(\bf{0}, \bfV)$ where the covariance matrix is diagonal $\bfV = diag(\sigma^2_1, \dots, \sigma^2_R)$, and the diagonal elements $\sigma^2_1, \dots, \sigma^2_{R}$ are known as  idiosyncratic variances. Note that Equation (\ref{MainEQ1}) encodes a dimension reduction from dimension $R$ to dimension $F$. 
To ensure the identifiability of the model, we assume that the matrix of factor loadings $\bfB$ follows a hierarchical structural constraint \citep{geweke:zhou:1996, aguilar:west:2000, prad:ferr:west:2021}. Specifically, the matrix of factor loadings $\bfB$ is assumed to be lower triangular with all main diagonal elements equal to 1, that is,
\begin{equation}
\bfB = 
\begin{bmatrix}
1 & 0 & 0 & \dots & 0\\
b_{2,1} & 1 & 0 & \dots & 0\\
b_{3,1} & b_{3,2} & 1 & \dots & 0\\
\vdots & \vdots & \vdots & \ddots & \vdots\\ 
b_{F,1} & b_{F,2} & b_{F,3} & \dots & 1\\
b_{F+1,1} & b_{F+1,2} & b_{F+1,3} & \dots & b_{F+1,F}\\
\vdots & \vdots & \vdots & \ddots & \vdots\\
b_{R,1} & b_{R,2} & b_{R,3} & \dots & b_{R,F}
\end{bmatrix}. \nonumber
\end{equation}
Note that each row of $\bfB$ corresponds to an observed variable, and each column corresponds to a common factor. The order of the variables should be carefully chosen; the current literature provides recommendations \citep{aguilar:west:2000, prad:ferr:west:2021,shin2023dynamic}.

BCFM assumes that the common factors $\bfx_1, \ldots, \bfx_n$ follow a mixture of Gaussian distributions with $K$ components \citep{west1992,escobar1995bayesian}. Each component of the mixture corresponds to a cluster, hence each subject belongs to one of $K$ clusters. Let $z_{i}$ indicate the cluster subject $i$ belongs to. Then, given $z_i = k$, BCFM assumes that the conditional distribution of the common factor~$\bfx_i$ is 
\begin{equation}\label{MainEQ2}
\bfx_{i} | z_{i} = k \sim N (\bfmu_k, \bfOmega_k),
\end{equation}
where $\bfmu_k = (\mu_{k1}, \dots, \mu_{kF})'$ is the mean vector and $\bfOmega_k$ is the covariance matrix of the common factors for subjects that belong to cluster $k$. To make the model identifiable, we impose a constraint that the covariance matrix $\bfOmega_1$ of the first cluster is diagonal. 
Let the probability of a randomly selected subject belonging to cluster $k$ be $p_k = P(z_{i} = k)$. Then, the Gaussian mixture model for $\bfx_i$ is
\begin{equation}\label{MainEQ3}
\bfx_{i} \sim  \sum_{k=1}^{K} p_k N (\bfmu_k, \bfOmega_k).
\end{equation}


\subsection{Priors}\label{Priors}

In this section, we present the priors that we use for data analysis with BCFM. We favor conditionally conjugate priors that allow the incorporation of a wide range of prior information and at the same time lead to straightforward computations \citep{prad:ferr:west:2021}. 

First, let us consider the prior for the elements of the factor loadings matrix $\bfB$. We assume that {\it a priori} the unconstrained elements in the $l$th column of $\bfB$, corresponding to the $l$th factor, are independent and identically distributed with a Gaussian distribution with mean 0 and variance $\tau_l$, $l = 1, \dots, F$. Thus, for $r > F$, the elements of the $r$th row of $\bfB$ follow {\it a priori} a Gaussian distribution with mean vector $\bfzero$ and covariance matrix  $\bfT = diag(\tau_1, \dots, \tau_F)$. In addition, for $2 \leq r \leq F$, the unconstrained elements of the $r$th row of $\bfB$ follow {\it a priori} a Gaussian distribution with mean vector $\bfzero$ and covariance matrix $diag(\tau_1, \dots, \tau_{r-1})$. 

For the idiosyncratic variances and the variances of the factor loadings, we assume conditionally conjugate inverse gamma priors. Specifically, for the variance $\tau_l$ of the factor loadings of the $l$th factor, $l=1, \ldots, F$, we assume an inverse gamma prior $IG(n_\tau/2, n_\tau s^2_\tau/2)$, where we follow the recommendations of  \cite{prad:ferr:west:2021} and choose $n_\tau$ = 1 and $s^2_\tau = 1$. In addition, for the  idiosyncratic variance $\sigma^2_r$ of the $r$th variable, we assume an inverse gamma prior $IG(n_\sigma/2, n_\sigma s_\sigma^2/2)$, $r=1, \ldots, R$. Specifically, following the recommendations of  \cite{lopes:west:2004} and \cite{prad:ferr:west:2021}, we choose $n_\sigma$ = 2.2 and $n_\sigma s^2_\sigma$ = 0.1. This choice of hyperparameters implies the idiosyncratic variances have a prior mean equal to 0.5 and infinite prior variance, which implies a vague prior distribution. 

For the vector of cluster probabilities $\bfp = (p_1, \dots, p_K)$ we assign a conditionally conjugate Dirichlet prior $Dirichlet(\alpha_1, \dots, \alpha_K)$. 
The choice of hyperparameters for this prior has to be done in a careful manner. In particular, while usual non-informative priors for probabilities assign  $(\alpha_1, \dots, \alpha_K) = (0.5,  \dots, 0.5)$ or $(\alpha_1, \dots, \alpha_K) = (1,  \dots, 1)$, for mixture models these prior choices allow clusters with very low probability that may become empty during the MCMC algorithm. Thus, to keep the probabilities $p_1, \dots, p_K$ away from zero, we assume prior $(\alpha_1, \dots, \alpha_K) = (2,  \dots, 2)$.

For the mean vector $\bfmu_k$ of the $k$th cluster we assign the Gaussian prior $\bfmu_k \sim N(\bfm_k, \bfC_k)$, where $\bfm_k$ and $\bfC_k$ are assumed to be known, $k = 1, \dots K$. 
Further, for the $l$th element of the diagonal covariance matrix $\bfOmega_1$ of the first cluster, we assign the inverse gamma prior $\{\Omega_{1}\}_{ll} \sim IG(n_{\omega_l}/2, n_{\omega_l} s^2_{\omega l}/2)$. Finally, for covariance matrix $\bfOmega_k$, we assign the conditionally conjugate inverse Wishart prior $\bfOmega_k \sim IW(\nu + F, \bfPsi_k)$, $k = 2, \dots, K$. 

Next, we propose a way to choose the prior hyperparameters for the mean vectors and covariance matrices of the clusters. Different approaches for data-dependent choice of hyperparameters in mixture models have been proposed \cite{raftery1996hypothesis,wasserman2000asymptotic,richardson1997bayesian,steele2010performance}. In particular, Steele and Raftery \cite{steele2010performance} developed unit information priors \citep{kass:1995} for mixtures of Gaussian distributions as models for observed data. Similarly to the work of Steele and Raftery \cite{steele2010performance}, our proposal does not assume the same prior for the mean vector and covariance matrix of each mixture component. As a result, this prior choice eliminates the issue of multimodality of the posterior distribution, and consequently ameliorates problems with MCMC convergence and label switching. However, in contrast to the proposal of Steele and Raftery \cite{steele2010performance} for mixtures of Gaussian distributions as models for observed data, in our work we use mixtures for latent factors which makes the choice of prior hyperparameters a somehow more difficult problem. 

To choose the prior hyperparameters for the cluster parameters, we propose an empirical Bayes approach that assigns weakly informative priors for the cluster mean vector $\bfmu_k$ and covariance matrix~$\bfOmega_k$, $k=1,\ldots,K$. Specifically, we determine the hyperparameters for these priors with a preliminary factor model analysis followed by $k$-means clustering. We implement this preliminary factor model analysis using the function {\it fa} from the R package {\it psych} \citep{psych:2025}. After running the {\it fa} function, let $\widehat{\bfB}$ be the estimated $R \times F$ matrix of factor loadings, $\widehat{\sigma}^2_1, \ldots, \widehat{\sigma}^2_R$ be the estimated idiosynchratic variances, and  $\widehat{\bfV}=\diag(\widehat{\sigma}^2_1, \ldots, \widehat{\sigma}^2_R)$. Further, let $\bfM$ be a matrix such that $\widehat{\bfB}^* = \widehat{\bfB} \bfM$ satisfies the hierarchical structural constraint. Then, we obtain a preliminary estimate of the vector of common factors for subject $i$ with
\begin{equation}
\widehat{\bfx}_i = (\widehat{\bfB}^{*'}  \widehat{\bfV}^{-1} \widehat{\bfB}^*)^{-1} \widehat{\bfB}^{*'}  \widehat{\bfV}^{-1} \bfy_i.
\end{equation}
We use the preliminary estimates of the vectors of common factors $\widehat{\bfx}_1, \dots, \widehat{\bfx}_n$ to determine the prior hyperparameters for $\bfmu$ and $\bfOmega$. 

Next, we apply $k$-means clustering to $\widehat{\bfx}_1, \dots, \widehat{\bfx}_n$. 
Let $\bfS_1, \dots, \bfS_K$ be the sample covariance matrices of the preliminary estimates of the vectors of common factors $\widehat{\bfx}_i$ within each of the $K$ clusters identified by the $k$-means clustering algorithm. Because $k$-means clustering may return a local optimum, we run the algorithm 50 times and choose the $k$-means solution that minimizes the sum of the Euclidean distances between each $\widehat{\bfx}_i$ and the corresponding cluster center. 

To ensure identifiability, BCFM assumes that the covariance matrix of the vector of common factors that belong to the first cluster is diagonal. To obtain a prior that respects this identifiability assumption, consider the LDL decomposition $\bfS_1 = \bfL_1 \bfD_1 \bfL_1'$ where $\bfL_1$ is a lower triangular matrix with diagonal elements equal to 1 and $\bfD_1$ is a diagonal matrix. In addition, let $\widetilde{\bfx}_i = \bfL_1^{-1} \widehat{\bfx}_i$. Then, the sample covariance matrix of the transformed preliminary estimate $\widetilde{\bfx}_i$ that belongs to the first cluster is $\bfL_1^{-1} \bfS_1 (\bfL_1^{-1})' = \bfD_1$, which is a diagonal matrix. Further, note that $\widetilde{\bfB} = \widehat{\bfB}^* \bfL_1$ also satisfies the hierarchical structural constraint and $\widetilde{\bfB} \widetilde{\bfx_i} = \widehat{\bfB} \widehat{\bfx}_i$. Therefore, both $\widetilde{\bfB}$ and $\bfD_1$ satisfy the BCFM identifiability constraints.

Let $\widetilde{Z}_k$ be the set of observations assigned by $k$-means to the $k$th cluster. Let $n_k$ be the number of observations in $\widetilde{Z}_k$. Then, for the mean vector of the $k$th cluster we assign the prior $\bfmu_k \sim N(\bfm_k,\bfC_k)$, where $\bfm_k = n_k^{-1} \sum_{i\in Z_k} \widetilde{\bfx}_i$ and $\bfC_k = \bfL_1^{-1} \bfS_k (\bfL_1^{-1})'$. 
In addition, the hyperparameters in the prior for the $l$th diagonal element of $\bfOmega_1$ are $n_{\omega_l} = 4$ and $s_{\omega_l}^2 = \{ \bfD_1 \}_{ll}$, which imply a prior mean equal to $s_{\omega_l}^2$. Further, for the covariance matrix of the $k$th cluster we assign the prior $\bfOmega_k \sim IW(\nu,\bfPsi_k)$, where $\nu = F + 2$ and $\bfPsi_k = \bfL_1^{-1} \bfS_k (\bfL_1^{-1})'$. This implies a prior mean for $\bfOmega_k$ equal to $\bfPsi_k$. 

The simulation study in Section~\ref{Simulation_Study} shows that these weakly informative priors work well, allowing the inferential approach proposed in the next section to provide adequate quantification of uncertainty. 

\section{Statistical Inference}\label{sec:Statistical_Inference}


\subsection{Posterior Exploration}\label{sec:Posterior_Exploration}

We propose an MCMC algorithm \citep{robe:case:2005, game:lope:2006} to explore the posterior distribution of the BCFM parameters. Specifically, this MCMC algorithm is a Gibbs sampler \citep{gelfand:smith:1990} that simulates draws from the full conditional distributions of the parameters. In this section, we present these full conditional distributions.

The full conditional distribution of the common factor $\bfx_i$ depends on the cluster assignment $z_i$. Given $z_i = k$, the full conditional of $\bfx_i$ is $N (\bfm_i, \bfA_i), i = 1, \dots, n$, where
\begin{eqnarray}\label{X_EQ}
\bfA_i &=& \left( \bfOmega_k^{-1} + \bfB' \bfV^{-1} \bfB \right)^{-1},  \\
\bfm_i &=& \left( \bfOmega_k^{-1} + \bfB' \bfV^{-1} \bfB \right)^{-1} \left(\bfB' \bfV^{-1} \bfy_{i} + \bfOmega^{-1}_k \bfmu_k \right).
\end{eqnarray}

The mean vector $\bfmu_k$ of the $k$th cluster, $k = 1, \dots, K$, has the following Gaussian full conditional distribution
\begin{equation}\label{mu_EQ}
\bfmu_k | \bfY, \bfX \sim N \left(\bfC_k^* \left( \bfC_k^{-1} \bfm_k + n_k \bfOmega_k^{-1} \bar{\bfX}^*_k \right), \bfC_k^* \right),
\end{equation}
where $\bfC_k^*=(\bfC^{-1}_k + n_k \bfOmega_k^{-1})^{-1}$ is the covariance matrix of this full conditional distribution, $\bar{\bfX}^*_k$ is the mean of common factors for the observations that belong to the $k$th cluster, and $n_k$ is the number of observations in the $k$th cluster. 

Recall that the first cluster covariance matrix $\bfOmega_1$ is diagonal. In particular, its $l$th diagonal element $\omega_{1l}$ has the following inverse gamma full conditional distribution
\begin{equation}\label{Omega_EQ1}
    \omega_{1l} | \bfY, \bfX \sim IG \left( \frac{1}{2}(n_1 + n_{\omega_l}), \frac{1}{2} \left( \sum_{i\in C_1} (x_{il} - \mu_{1l})^2 + n_{\omega_l} s^2_{\omega_l} \right) \right),
\end{equation}
where $x_{il}$ is the $l$th element of $\bfx_i$, $\mu_{1l}$ is the $l$th element of $\bfmu_1$, $C_1$ is the set of observations that belong to the first cluster, and $n_1$ is the number of observations in $C_1$.

The covariance matrix $\bfOmega_k$ of the $k$th cluster, $k = 2, \dots, K$, has the following inverse Wishart full conditional distribution 
\begin{equation}\label{Omega_EQ2}
\bfOmega_k | \bfY, \bfX \sim IW\left( n_k + \nu, \sum_{i \in C_k}(\bfx_{i} - \bfmu_k ) (\bfx_{i} - \bfmu_k)' + \bfPsi_k \right),
\end{equation}
where $C_k$ is the set of observations that belong to the $k$th cluster and $n_k$ is the number of observations in $C_k$.

To simulate the matrix of factor loadings $\bfB$, we consider two cases: when $r > F$, and $1 < r \leq  F$. Let $\bfB_r$ be the $r$th row of $\bfB$. For $r > F$, the full conditional distribution of $\bfB_r$ is the multivariate Gaussian distribution 
\begin{equation}\label{B_EQ1}
\bfB_r | \bfY, \bfX \sim N \left( \frac{1}{\sigma^2_r} \bfT^*_r  \bfX' \bfy_{.,r} , \bfT^*_r\right),
\end{equation}
where 
$\bfT^*_r=( \sigma^{-2}_r \bfX' \bfX + \bfT^{-1})^{-1}$ is the covariance matrix of this full conditional distribution, $\bfy_{.,r}$ is the $r$th column of $\bfY$, and $\bfT = diag(\tau_1, \dots, \tau_F)$ is the prior covariance matrix of each row of $\bfB$. Now let us consider the case when $1 < r \leq F$. Due to the hierarchical structural constraint, the last $F-r+1$ elements of $\bfB_r$ are fixed. Thus, for $1 < r \leq F$, there are $r - 1$ free elements in $\bfB_r$, and no free elements when $r = 1$. Let $\bfX_{. , 1:(r-1)}$ be submatrix of $\bfX$ before the $r$th column and $\bfX_{. , r}$ be the $r$th column of $\bfX$. Also, let $\bfT_{1:(r-1), 1:(r-1)}$ be the submatrix of the first $r-1$ rows and columns of the factor loading covariance matrix $\bfT$. Then, the full conditional of $\bfB_r$ is the Gaussian distribution $\bfB_r \sim N(\bfQ_r \bfa_r, \bfQ_r)$ where
\begin{eqnarray}\label{B_EQ2}
\bfQ_r &=& \left(\frac{1}{\sigma^2_r} \left(\bfX_{. , 1:(r-1)}' \bfX_{. , 1:(r-1)} \right) + \bfT_{1:(r-1), 1:(r-1)} \right)^{-1}, \\
\bfa_r &=& \frac{1}{\sigma^2_r} \left( \bfX'_{.,1:(r-1)} (\bfy_{. , r} - \bfX_{. , r}) \right). 
\end{eqnarray}

The full conditional distribution of the $r$th idiosyncratic variance $\sigma^2_r, r = 1, \dots, R,$ is the inverse gamma distribution
\begin{equation}\label{sigma2_EQ}
\sigma^2_r | \bfY, \bfX \sim IG \left(\frac{1}{2}(n_\sigma + S), \frac{1}{2} \left(n_\sigma s^2_\sigma + \sum_{i=1}^n (\bfY_{ir} - \bfB_{r}\bfx_{i})^2 \right) \right).
\end{equation}

The full conditional distribution of the variance among factor loadings of the $l$th factor $\tau_l$, $l = 1, \dots F$, is an inverse gamma distribution. Recall that the first $l$ elements of the $l$th factor are fixed at 0 or 1 by the hierarchical structural constraint. Let $\bfB_{(l+1:r),l}$ be the $l$th factor after the first $l$ elements. Then, the full conditional distribution of $\tau_l$ is
\begin{equation}\label{tau_EQ}
\tau_l | \bfB_{(l+1):r,l} \sim IG \left(\frac{1}{2}(R - l + n_\tau), \frac{1}{2}(\bfB'_{(l+1):r,l} \bfB_{(l+1):r,l} + n_\tau s_\tau^2) \right).
\end{equation}

The full conditional distribution of the cluster assignment of the $i$th subject $z_i$ is the discrete distribution 
\begin{equation}\label{Z_EQ}
p(z_{i} = k | \bfY, \bfX) \propto p_k|\bfOmega_k|^{-1/2}exp \left(-\frac{1}{2} (\bfx_{i} - \bfmu_k)' \bfOmega_k^{-1} (\bfx_{i} - \bfmu_k) \right). 
\end{equation}

The full conditional distribution of the vector of cluster probabilities $(p_1, \dots, p_K)$ is the Dirichlet distribution
\begin{equation}\label{probs_EQ}
p_1, \dots, p_K | \bfY, \bfX \sim Dirichlet ( n_1 + \alpha_1, \dots, n_K + \alpha_K ).
\end{equation}

Using the full conditional distributions presented above, here is the Gibbs sampler to explore the posterior distribution of the unknown quantities from the BCFM:
\begin{enumerate}
    \item Set initial values for $\bfX$, $\bfB$, $\sigma_1^2, \dots, \sigma_R^2$, $\tau_1, \dots, \tau_F$, $\bfmu_1, \ldots, \bfmu_K$, $\bfOmega_1, \ldots, \bfOmega_K$, $z_1, \dots, z_n$, and $\bfp$. 
    \item Simulate $\bfX$ from the Gaussian distribution given in Equation (\ref{X_EQ}).
    \item Simulate $\bfmu$ from the Gaussian distribution given in Equation (\ref{mu_EQ}).
    \item Simulate each $\omega_{1l}, \dots \omega_{1F}$ from the inverse gamma distribution given in Equation (\ref{Omega_EQ1}) and $\bfOmega_2, \dots, \bfOmega_K$ from the inverse Wishart distribution given in (\ref{Omega_EQ2}).
    \item Simulate $\bfB$ from the Gaussian distributions given in Equations (\ref{B_EQ1}) and (\ref{B_EQ2}).
    \item Simulate each $\sigma_1^2, \dots, \sigma^2_R$ from the inverse gamma distribution given in Equation (\ref{sigma2_EQ}).
    \item Simulate each $\tau_1, \dots \tau_F$ from the inverse gamma distribution given in Equation (\ref{tau_EQ}).
    \item Simulate each $z_1, \dots, z_n$ from the discrete distribution given in Equation (\ref{Z_EQ}). 
    \item Simulate $\bfp$ from the Dirichlet distribution given in Equation (\ref{probs_EQ}). 
    \item Repeat Steps (2) to (9) until the MCMC algorithm converges and we have enough posterior draws.
\end{enumerate}

\subsection{Model Selection Information Criterion}\label{Model_Selection}

In practice, the number of clusters and the number of factors are not known. In this section, we propose an information criterion for model selection to choose the number of clusters and the number of factors. 

The information criterion we propose is similar to the Bayesian Information Criterion (BIC). This is related to the use of the BIC for choice of the number of components in a mixture of Gaussian distributions as developed by Roeder and Wasserman \citep{roeder1997practical}. While the original BIC considers the maximum likelihood estimates of the parameters, here we follow Roeder and Wasserman \citep{roeder1997practical} and use the posterior means. Thus, the information criterion we propose is defined as 
\begin{equation}\label{eq:BIC_like}
IC = d \log(n) - 2 \log p(\bfy|K, F, \widehat\bftheta),
\end{equation}
where 
$\widehat\bftheta=(\widehat{\bfB}, \widehat{\bfmu}, \widehat{\bfOmega}, \widehat{\bftau}, \widehat{\bfsigma}^2, \widehat{\bfp})$, and  
$\widehat{\bfB}, \widehat{\bfmu}, \widehat{\bfOmega}, \widehat{\bftau}, \widehat{\bfsigma}^2$ and $\widehat{\bfp}$ are the posterior means of $\bfB, \bfmu, \bfOmega, \bftau, \bfsigma^2$ and $\bfp$ computed from the output of the MCMC algorithm.
In addition, $p(\bfy|K, F, \widehat\bftheta)$ is the integrated likelihood 
\begin{equation}\label{Integrated_Likelihood}
\begin{aligned}
p(\bfy | K, F, \widehat\bftheta) = & \prod_{i=1}^n \sum_{k = 1}^K \widehat{p}_{k} (2\pi)^{-R/2}|\widehat{\bfB} \widehat{\bfOmega}_k \widehat{\bfB}' + \widehat{\bfV}|^{-1/2} \\ 
& exp \left(-\frac{1}{2}(\bfy_{i} - \widehat{\bfB}\widehat{\bfmu}_k)'(\widehat{\bfB}\widehat{\bfOmega}_k \widehat{\bfB}' + \widehat{\bfV})^{-1} (\bfy_{i} - \widehat{\bfB}\widehat{\bfmu}_k)\right),
\end{aligned}
\end{equation}
that is obtained by integrating out the latent factors and the cluster assignment variables. Finally, $d$ is the number of unknown parameters in $(\bfB,\bfmu, \bfOmega, \bftau,\bfsigma^2, \bfz, \bfp)$ which, in the case of BCFMs, is equal to
$$d = \frac{(K - 2) (F + 1)}{2} + (R + K) (F + 1) + F - 1.$$ 

Models with smaller information criterion are preferable. While our information criterion works well for most datasets, for a small number of datasets the information criterion may point to a model with a cluster that is empty or has a small number of observations. This may be related to results obtained by \cite{rousseau2011asymptotic} on the asymptotic behavior of the posterior distribution for a mixture model that has a number of clusters larger than the true number of clusters, with the caveat that their results may not be directly applicable to BCFM because of our distinct prior distributions for the cluster parameters. To ameliorate the issue of clusters with small number of observations, we regard a model as not acceptable and assign to it an information criterion $IC=\infty$ if one of its clusters has been inferred to have a small number of observations. As the simulation study reported in Section~\ref{Simulation_Study_for_Model_Selection} shows,  our proposed information criterion works well at guiding the selection of the number of factors and the number of clusters. 

Other Bayesian model selection criteria could be developed for BCFMs. For example,  
Lopes and West \citep{lopes:west:2004} provide a comparison of the performance of several Bayesian criteria for the choice of the number of factors in factor models and Steele and Raftery \citep{steele2010performance} compare the performance of several criteria for the choice of the number of components in Gaussian mixture models. In particular, Lopes and West \citep{lopes:west:2004} found that the marginal density approximated with the Laplace-Metropolis estimator  \citep{lewis:raft:1997} works well for the choice of the number of factors in factor models (\citep[see also Prado et al.][]{prad:ferr:west:2021}). We performed some preliminary explorations with the Laplace-Metropolis marginal density for BCFMs, but the information criterion we propose here performed much better.

\section{Simulation Studies}\label{Simulation_Study}

\subsection{Evaluation of Estimation}\label{Evaluation_of_Estimation}

To evaluate the quality of estimation, we consider a dataset simulated with $n = 1,000$ subjects, $R = 20$ variables, $K = 4$ clusters, and $F = 3$ factors. This simulation setting has been inspired by our recent analysis of individuals in recovery from opioid use disorder \citep{craft:shin:tegge:2022}. In our setting, the 1,000 subjects are randomly assigned to the 4 clusters with probabilities (0.45, 0.30, 0.15, 0.10).
The true mean vectors of the common factors of each cluster are $\bfmu_1 = (0.50, -0.50, 0.00)'$, $\bfmu_2 = (-1.50, -4.00, 2.50)'$, $\bfmu_3 = (-3.75, 2.50, 1.00)'$, and $\bfmu_4 = (-7.50, -1.75, 5.25)'$. The true values of the covariance matrices of the common factors of each cluster are 
\begin{equation*}
\bfOmega_1 = \left[
\begin{array}{ccc}
2.0 & 0.0 & 0.0 \\
0.0 & 1.0 & 0.0 \\
0.0 & 0.0 & 1.5 
\end{array}
\right], \ \ 
\bfOmega_2 = \left[
\begin{array}{ccc}
2.0 & 0.4 & 0.4 \\
0.4 & 2.0 & 0.4 \\
0.4 & 0.4 & 2.0 
\end{array}
\right],
\end{equation*}
\begin{equation*}
\bfOmega_3 = \left[
\begin{array}{ccc}
3.0 & 0.3 & 0.3 \\
0.3 & 3.0 & 0.3 \\
0.3 & 0.3 & 3.0 
\end{array}
\right], \text{ \ and \ }
\bfOmega_4 = \left[
\begin{array}{ccc}
4.0 & 1.0 & 1.0 \\
1.0 & 4.0 & 1.0 \\
1.0 & 1.0 & 4.0 
\end{array}
\right].
\end{equation*}
The true variances of the free elements of the matrix of factor loadings are $\tau_1 = 0.05$, $\tau_2 = 0.10$, and $\tau_3 = 0.15$. In addition, the idiosyncratic variances are all equal $\sigma^2_j = 0.1,$ for $j = 1, \dots, R$. 

To analyze the simulated dataset, first we assigned priors as explained in Section~\ref{Priors}. After that, we ran the MCMC algorithm proposed in Section~\ref{sec:Posterior_Exploration} for 50,000 iterations. To reduce computer memory burden we have kept one draw for each 10 iterations, retaining a total of 5,000 draws. Trace plots of the simulated quantities indicate that the algorithm converged after 1,500 draws (i.e., 15,000 iterations.) Thus, we discarded the first 1,500 draws as burn-in and used the remaining 3,500 draws to estimate the BCFM parameters. We performed posterior exploration for a total of 25 models, with number of clusters varying from 1 to 5 and number of factors varying from 1 to 5 as well. The information criterion we propose in Section~\ref{Model_Selection} selects the true model with 4 clusters and 3 factors. Henceforth, we present the results when fitting the model with $K = 4$ clusters and $F = 3$ factors.

Figure \ref{SD_Probability_Density} shows the posterior density of each of the cluster probabilities $\text{p}_1, \text{p}_2, \text{p}_3$, and $\text{p}_4$, as well as their respective true values (vertical dashed lines). The posterior modes are close to the true values and, thus, our BCFM framework is able to accurately estimate the cluster probabilities. In addition, for each cluster the true cluster probability falls within the respective 95\% credible interval. Therefore, our BCFM framework provides adequate quantification of uncertainty for the cluster probabilities.

Figure \ref{SD_mu_Density} displays the posterior densities of the elements of the cluster mean vectors $\bfmu_1, \dots, \bfmu_4$, where each of these vectors contains three elements, one for each factor. Panel (A) shows the posterior densities of the first element --- which corresponds to the mean of the first common factor --- of $\bfmu_1, \dots, \bfmu_4$. Panels (B) and (C) show analogous plots for the second and third elements, respectively. The modes of the posterior distributions are very close to the true values, the posterior densities are highly concentrated, and all the true values are located within the respective 95\% credible intervals. 

Figure \ref{SD_B_Density} shows the posterior mean and 95\% credible intervals of the factor loadings. True values are represented with blue triangles, posterior means are represented with black circles, and black vertical lines indicate 95\% credible intervals. 
Note that because of the hierarchical structural constraint, for the $l$th factor, the $l$th loading is fixed at 1 and the first to $(l-1)$th loadings are fixed at 0. The credible intervals are very narrow, and the posterior means visually overlap the true values. Specifically, 98.1\% of the true factor loadings are included in the 95\% credible intervals. 


Figure \ref{SD_sigma2_Density} shows true values (red dashed line), posterior means (black circles), and 95\% credible intervals (black vertical lines) for $\sigma^2_r$, $r = 1, \dots, 20$. Recall that the true values used to generate the simulated data are $\sigma^2_1 = \sigma^2_2 = \dots = \sigma^2_{20} = 0.1$. The posterior means are close to the true value indicating that our approach accurately estimates the idiosyncratic variances. In addition, the 95\% credible intervals include the true values 100\% of the time. 

Figure \ref{SD_heatmap} presents the heatmap of the posterior probability that each subject belongs to each cluster. The x-axis represents the 4 clusters, and the y-axis represents the 1,000 subjects. The blue lines separate the true clusters. Our BCFM framework correctly assigns 96\% of the subjects to their true clusters. Therefore, our approach assigns subjects to clusters with high accuracy.

In summary, the prior specification we propose in Section~\ref{Priors} combined with the MCMC approach we propose in Section~\ref{sec:Posterior_Exploration} lead to accurate estimation of the model parameters with appropriate quantification of uncertainty. 



\subsection{Simulation Study for Model Selection}\label{Simulation_Study_for_Model_Selection}


To further validate our BCFM framework, we perform a simulation study to compare the BCFM model selection approach with an approach that combines principal component analysis (PCA) and $k$-means clustering \citep[e.g., see][]{craft:shin:tegge:2022}.  Here, we compare BCFM and PCA+$k$-means in terms of the performance of correctly choosing the number of clusters and factors. For PCA+$k$-means, we use the Kaiser criterion \citep{kaiser:1960} that chooses the number of factors as equal to the number of eigenvalues larger than one, and we use the gap statistic \citep{tibshirani2001estimating} to choose the number of clusters. For BCFM, we choose the model with the smallest information criterion proposed in Section~\ref{Model_Selection}. 

We consider settings with ten different levels of separation between clusters. To implement the different levels of separation, the mean vectors of each cluster are obtained by multiplying fixed vectors by a scalar $s$ that varies from 0.1 to 1. When $s=0.1$ the cluster mean vectors are closer to each other whereas when $s=1$ the cluster mean vectors are farther away from each other. Thus, $s$ indicates the degree of separation of the clusters. The mean vectors for the $K = 4$ clusters under the setting with separation equal to $s$ are
\begin{eqnarray*}
\bfmu_1 & = & s \times (1.0, -1.0, 0.00)', \\
\bfmu_2 & = & s \times (-3.0, -8.0, 5.0)', \\
\bfmu_3 & = & s \times (-7.5, 5.0, 2.00)', \mbox{and} \\
\bfmu_4 & = & s \times (-15.0, -3.5, 10.5)'.
\end{eqnarray*}
For each separation $s$ in the set $\{0.1, 0.2, \ldots, 1.0\}$ we have simulated 100 datasets. In all considered settings, we simulate datasets with $K = 4$ clusters and $F = 3$ factors. All other parameters are the same as in the simulated example in Section~\ref{Evaluation_of_Estimation}. 
 
\begin{figure}
\centering
\includegraphics[width=8cm]{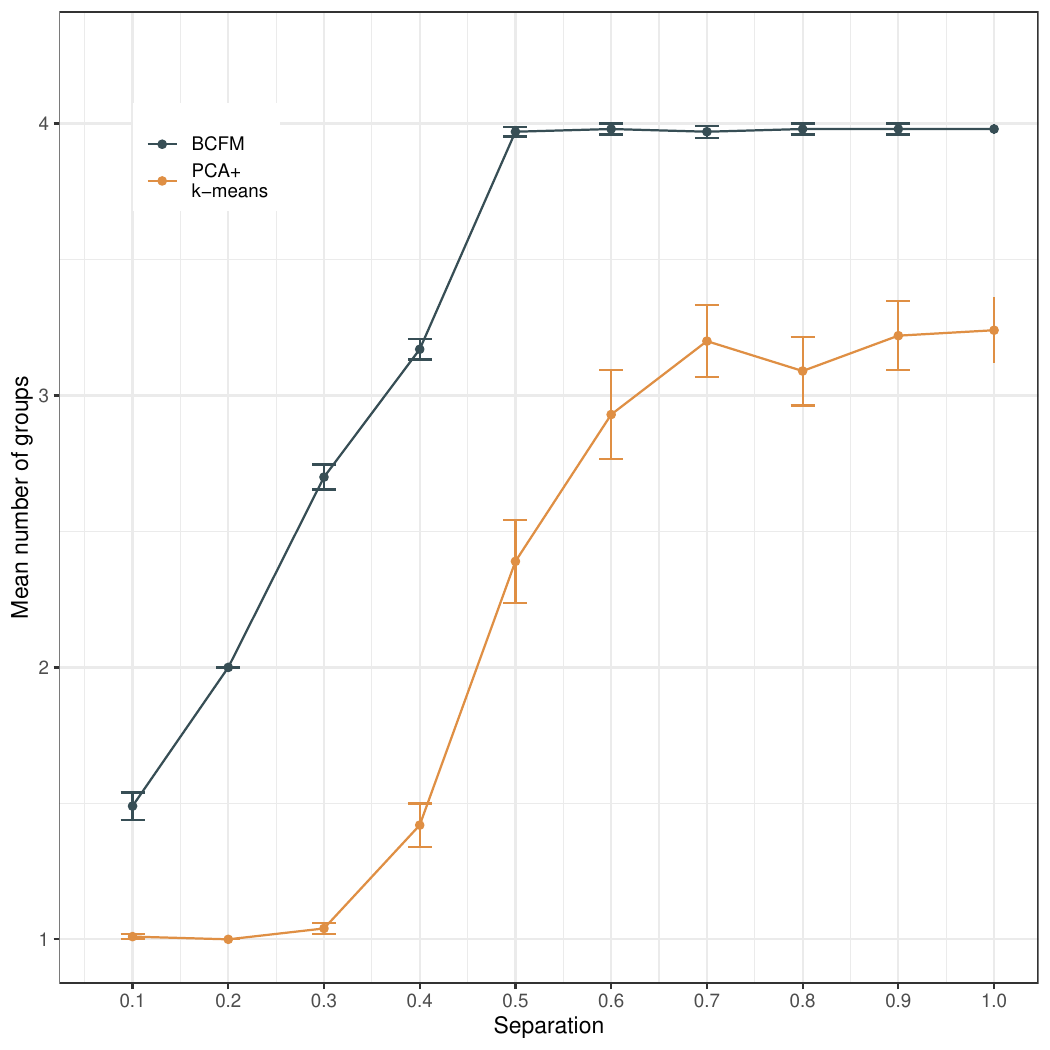}
\caption{BCFM versus PCA+$k$-means. Mean number of clusters (with standard error) selected by each method as a function of separation. The true number of clusters is 4.}
\label{fig:BCFM-versus-PCAKMEANS}
\end{figure}
Figure~\ref{fig:BCFM-versus-PCAKMEANS} presents the mean number of clusters (with standard error) selected by each method as a function of separation. Recall that the true number of clusters is $K=4$. Many features can be observed from this figure. First, the lower the separation among clusters, the harder it is to identify the true number of clusters. This is evidenced by separation $s=0.1$ where the PCA+$k$-means approach usually identifies only one cluster, whereas the BCFM approach identifies on average of 1.5 clusters. As the separation among clusters increases, the performance of BCFM at identifying the number of clusters improves. For separations larger or equal to 0.5, BCFM performs very well and identifies on average close to 4 clusters. In contrast, the PCA+$k$-means approach tends to improve at identifying the number of clusters  as the separation increases, but on average identifies a smaller number of clusters than the true number of clusters. For example, at separation $s=0.5$, while BCFM identifies on average close to the true number of clusters $K=4$, PCA+$k$-means identifies on average about 2.25 clusters. Even at separation $s=1.0$, PCA+$k$-means identifies on average about 3.25 clusters, while BCFM identifies on average about 4 clusters. Therefore, when compared to PCA+$k$-means, BCFM performs much better at identifying the number of clusters. 

The selection of the number of factors is somehow easier than the selection of the number of clusters. Case in point, considering all 1,000 simulated datasets, BCFM selected the correct number of factors $F=3$ for 996 datasets. In addition, PCA+$k$-means did reasonably well when selecting the number of factors for separations less or equal than 0.7, making just one mistake in those cases. However, the performance of PCA+$k$-means deteriorates when selecting the number of factors for separations larger or equal than 0.8. For example, for separation $s=1$, PCA+$k$-means selects the correct number of factors $F=3$ for only 79\% of the datasets, and for the remaining 21\% of the datasets it selects 2 factors. Therefore, when compared to PCA+$k$-means, BCFM performs much better at identifying the number of factors. 

In summary, BCFM performs much better than PCA+$k$-means at identifying both the number of clusters and the number of factors. 

\section{Application to Recovery from Opioid Use Disorder}\label{sec:OUD_Recovery_Data}

This section presents a BCFM analysis of a dataset on recovery from opioid use disorder. The data are from the Remission from Chronic Opioid Use -- Studying Environmental and SocioEconomic Factors on Recovery (RECOVER, NCT03604861) Study \citep{ling:vijay:2019}. While the RECOVER study collected data for 24 months, here we focus on the data from the first time point that is referred to as baseline. 

The dataset we consider has  $n = 348$ participants with complete data from $R = 13$ variables. The variables are the Subjective Opiate Withdrawal Scale (SOWS), Beck's Depression Inventory \RNum{2} (BDI), Family \& Social conflict scores, Brief Pain Inventory (BPI; 3-items representing average, worst, and least pain), Kessler's psychological distress (K6), one question about the need for lifetime opioid use disorder medication, one question about confidence in abstinence,  the physical and mental categories of the 12-Item Short Form Survey (SF-12), and one question related to interview quality.

To select the number of factors and clusters we compute the information criterion proposed in Section~\ref{Model_Selection} for models with number of factors varying from 1 to 5 and with number of clusters varying from 1 to 6. 
Table \ref{tab:BL_G4K3_BIC} shows the result our information criterion for the different combination of number of clusters and factors. According to this criterion, the 5-cluster and 4-factor model is the best model. Therefore, henceforth we present results for the model with 5 clusters and 4 factors.

\begin{figure}
\centering
\includegraphics[width=8cm]{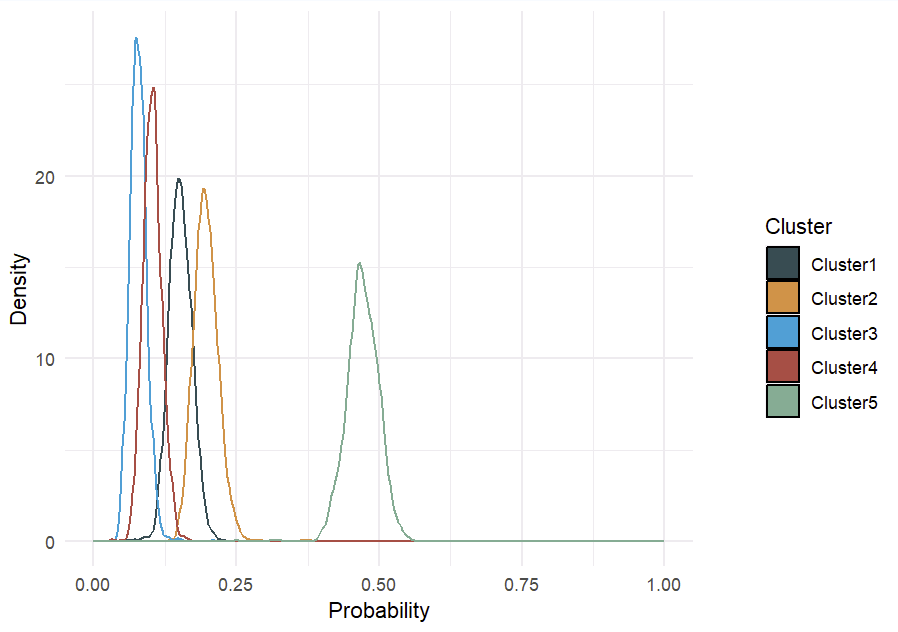}
\caption{Posterior densities of cluster assignment probabilities for the 5-cluster and 4-factor BCFM on the RECOVER data.}\label{BL_Probability_Density}
\end{figure}



 Figure \ref{BL_Probability_Density} shows the posterior density of the cluster assignment probabilities. Four of the clusters have cluster assignment probabilities less than 0.20 while the largest cluster has cluster assignment probability at 0.47. The posterior mean of the vector of assignment probabilities is (0.15, 0.20, 0.08 0.10, 0.47).

\begin{figure}
\centering
\includegraphics[width=8cm]{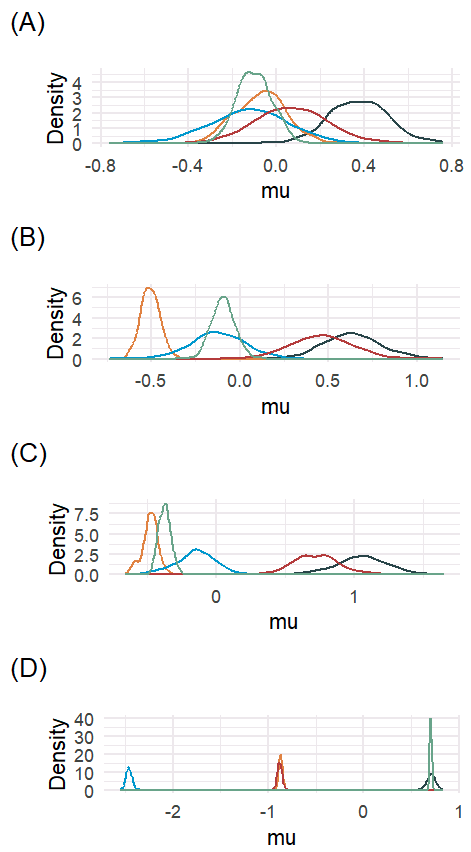}
\caption{Posterior density of the cluster means for the first common factor (Panel A), the second common factor (Panel B),  the third common factor (Panel C), and the fourth common factor (Panel D) on the RECOVER data. }\label{BL_mu_Density}
\end{figure}



Figure \ref{BL_mu_Density} shows the posterior densities of the cluster means for the first common factor (Panel A), the second common factor (Panel B),  the third common factor (Panel C), and the fourth common factor (Panel D). The uncertainty in the estimation of the cluster means is small when compared to the variation across the cluster means. In addition, this uncertainty is smaller for larger clusters. The first common factor separates cluster 1 from the other three clusters. The means of common factors 2 and 3 are similar for clusters 1 and 4. Finally, common factor 4 separates cluster 3, from clusters 2 and 4, and clusters 1 and 5.




Table \ref{tab:BL_G4K3_B_matrix} indicates the posterior means of the factor loadings matrix $B$. Note that the uncertainty in the estimation of the factor loadings matrix was similar to that observed in the simulation study. Average pain, BDI, family conflict, and confidence in abstinence were fixed to 1 for the first four factors, respectively. For factor 1, the variables with the largest factor loadings include the three pain variables followed by the physical health quality of life. For Factor 2, the variables will the largest factor loadings include BDI, K6, and mental quality of life. For Factor 3, the variables with the largest factor loadings include both family and social conflict. Finally for Factor 4, Confidence in abstinence has the largest factor loading followed by the need for lifetime medication for opioid use disorder. Finally, note that interview quality, which was included as a negative control, did not have strong loadings on any factor.



Figure \ref{BL_sigma2_Density} displays the posterior density of the idiosyncratic variances $\sigma_1^2, \dots, \sigma^2_{13}$. Idiosyncratic error variance of the first, fourth, and ninth variables are close to zero. The 95\% credible intervals are also narrower than the other variables. The variables with the largest posterior mean are the family conflict (varibales 3), interview question (variable 12), and need for lifetime medication for opioid use disorder (variable 13), with values at 0.96, 0.95, and 0.91, respectively. 

\begin{figure}
\centering
\includegraphics[width=.45\textwidth]{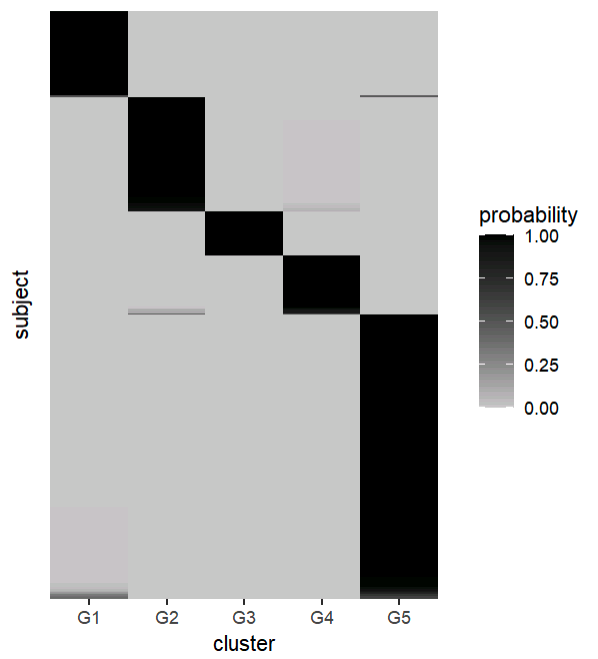}
\caption{Heatmap of the cluster assignments on the RECOVER data. The subjects are ordered according to the largest cluster probabilities.}\label{BL_Heatmap}
\end{figure}



Figure \ref{BL_Heatmap} is the heatmap of the cluster assignments. Overall, there is little variability when assigning individuals to clusters. However, we observe some overlap between cluster 1 and cluster 5, as well as cluster 2 and cluster 4.

\section{Conclusions}\label{sec:Conclusions}

We have proposed BCFM, which combines Gaussian mixture models and factor models for concomitant dimension reduction and clustering. In addition, we have developed a Gibbs sampler for estimation of the parameters of the model. Finally, we have proposed an information criterion for joint selection of the number of factors and the number of clusters. 

Simulation studies show that our BCFM framework accurately estimates the model parameters. In addition, we compared the performance of our proposed BCFM information criterion for selecting the number of factors and number of clusters versus the Kaiser criterion and the gap statistics combined with PCA+$k$-means. 
Overall, BCFM correctly identified the number of clusters for moderately (separation=0.5) to well-separated (separation=1.0) clusters. Meanwhile, the gap statistics combined with PCA+$k$-means consistently underestimated the number of clusters regardless of separation. 

We illustrated the real-world usability of BCFM on an opioid use disorder dataset. In the RECOVER dataset, BCFM information criterion selected 5 clusters and 4 factors as the best model. In this study, most subjects had high posterior probability of assignment to one cluster, with a few participants showing some uncertainty in cluster assignment.

There are many possible avenues for future research. One such avenue that we are currently exploring is the extension of BCFM to longitudinal studies where multivariate data for many subjects is collected over time. This would be useful for longitudinal studies of opioid use disorder recovery~\citep{craft:shin:tegge:2022}, and would enable researchers to identify subject-specific temporal trajectories through clusters. 
Another promising subject for future research would be the extension of BCFM to mixed-type data. This would be useful for the analysis of datasets where each subject provides multivariate data with a mix of binary, count, and continuous variables.

\section*{Acknowledgments}
This research was funded by a grant from the Academy of Data Science Discovery Fund at Virginia Tech and by award R21DA057580 from the National Institutes of Health.

\bibliographystyle{wileyNJD-AMS} 
\bibliography{BCFM}

\clearpage

\begin{table}
\begin{center}
\begin{tabular}{|l|l|l|l|l|l|l|}
\hline
\textbf{Model} &  K = 1 & K = 2 & K = 3 & K = 4 & K = 5 & K = 6 \\ \hline
F = 1 & 11700 &  Inf &  Inf &  Inf &  Inf & Inf \\ \hline
F = 2 & 11178 & 10794 & 10711 & 10691 & 10611 & 10610 \\ \hline
F = 3 & 11186 & 10248 & 9983 & 10642 &  Inf & Inf \\ \hline
F = 4 & 11191 & 10078 & 9762 & 9665 & \textbf{9128} & Inf \\ \hline
F = 5 & 11212 & 10100 & 10077 & 9654 & Inf & Inf \\ \hline
\end{tabular}
\vspace{.2 in}
\caption{\label{tab:BL_G4K3_BIC}  Information criterion for BCFM on the RECOVER data.}
\end{center}
\end{table}

\begin{table}
\begin{center}
\begin{tabular}{|l|c|c|c|c|}
\hline
{\bf Variable} & {\bf Factor 1} & {\bf Factor 2} & {\bf Factor 3} & {\bf Factor 4} \\ \hline
Pain -- average         &     1   &    0    &   0   &    0  \\  \hline
 BDI              &       0.29  &  1    &   0   &    0  \\  \hline
 Family conflict     &    0.13  &  -0.15  & 1    &   0  \\  \hline
Confidence in abstinence  & -0.01 &  0.03 &   -0.01  & 1  \\  \hline
Pain -- worst         &     0.89  &  0   &    0.04   & -0.05\\\hline
Pain -- best           &    0.83  &  0.01  &  -0.02 &  0.02 \\\hline
 SOWS                  &  0.26  &  0.37  &  0.01  &  0.01 \\\hline
 K6                    &  0.23 &   0.89  &  0.05 &   -0.06\\\hline
 Social conflict       &  0.11  &  -0.01  & 1.08  &  -0.07\\\hline
Mental QOL    & -0.28  & -0.78  & -0.07 &  0.04 \\\hline
Physical health QOL              &  -0.56 &  -0.11  & -0.09 &  -0.03\\\hline
Interview quality            &   0.02  &  0.08   & -0.08 &  0.21 \\\hline
Need for lifetime MOUD         &  -0.04  & -0.03 &  0.03  &  -0.3 \\ \hline
\end{tabular}
\vspace{.2 in}
\caption{\label{tab:BL_G4K3_B_matrix} Estimated factor loadings  matrix $B$ for BCFM on the RECOVER data. BDI: Beck depression index. SOWS: subjective opioid withdrawl scale. K6: Kessler's psychological distress scape. QOL: quality of life. MOUD: medication for opioid use disorder.}
\end{center}
\end{table}

\onecolumn
\section{Appendix}\label{Appendix}

\subsection*{Full conditional distribution of factor loadings matrix $\bfB$}

Consider $\bfb_r$ be the (column) vector that contains the $r$th row of the factor loadings matrix $\bfB$. Recall that $\bfT = diag(\tau_1, \dots, \tau_F)$ is the prior covariance matrix of each row of $\bfB$. 

When $r=1$, due to the hierarchical structural constraint, the first row is fixed $\bfb_1=\bfb_r=(1,0,\ldots,0)$.

When $r > F$, the full conditional density of $\bfb_r$ is
\begin{align*}
    p(\bfb_r| -)&\propto f(\bfY | \bfX, \bfB, \bfV) p(\bfb_r)\\
    &\propto exp \left( -\frac{1}{2} \sum_{i=1}^n (\bfy_{i} - \bfB \bfx_{i})' \bfV^{-1} (\bfy_{i} - \bfB \bfx_{i}) \right) exp \left( -\frac{1}{2} \bfb_r' \bfT^{-1} \bfb_r \right)\\
    &\propto exp \left( -\frac{1}{2 \sigma^2_r} (\bfy_{.,r} - \bfX \bfb_r)' (\bfy_{.,r} - \bfX \bfb_r) \right) exp \left( -\frac{1}{2} \bfb_r' \bfT^{-1} \bfb_r \right)\\
    &\propto exp \left( -\frac{1}{2} \left( \bfb_r' \left( \frac{1}{\sigma^2_r} \bfX' \bfX + \bfT^{-1} \right) \bfb_r -\frac{2}{\sigma^2_r} \bfb_r' \bfX' \bfy_{.,r} \right) \right).
\end{align*}

Therefore, when $r > F$, the full conditional distribution of $\bfb_r$ is $N(( \sigma^{-2}_{r}\bfX' \bfX + \bfT^{-1} )^{-1} \bfX' \bfy_{.,r},$ $( \sigma^{-2}_{r}\bfX' \bfX + \bfT^{-1} )^{-1})$.

Now consider the case when $1 < r \leq F$. Due to the hierarchical structural constraint, the last $F - r + 1$ elements of $\bfb_r$ are fixed, where  the $r$th element is equal to 1 and the last $F - r$ elements are equal to 0. Let $\bfb_r^*$ be the vector that contains the first $r-1$ elements of  $\bfb_r$. Then, the full conditional distribution of $\bfb_r^*$ is
\begin{align*}
    p(\bfb_r^*| -) &\propto f(\bfY| \bfX, \bfB, \bfV) p(\bfb_r^*)\\
    &\propto exp \left( -\frac{1}{2 \sigma^2_r}  (\bfy_{.,r}' - \bfX'_{.,r} - \bfb^{*'}_{r} \bfX'_{.,1:r-1}) (\bfy_{.,r} - \bfX_{.,r} - \bfX_{.,1:r-1} \bfb^{*}_{r}) \right) exp \left( -\frac{1}{2} \bfb_{r}^{*'} \bfT_{1:r-1,1:r-1}^{*-1} \bfb_{r}^* \right)\\
    &\propto exp \left( -\frac{1}{2} \left( \bfb_r^{*'} \left( \frac{1}{\sigma^2_r} \bfX'_{.,1:r-1} \bfX_{1:r-1} + \bfT_{1:r-1,1:r-1}^{-1} \right) \bfb^*_r -2 \frac{1}{\sigma^2_r} \bfb_r^{*'} \bfX'_{.,1:r} (\bfy_{.,r} - \bfX_{.,r}) \right) \right).
\end{align*}

Therefore, when $1 < r \leq F$, the full conditional distribution of $\bfb_r^*$ is  $N(( \sigma^{-2}_{r} \bfX_{.,r}' \bfX_{.,r} + \bfT_{1:r-1,1:r-1}^{-1} )^{-1} \bfX_{.,1:r-1}'$ 
$(\bfy_{.,r} - \bfX_{.,r}) / \sigma^2_r, ( \sigma^{-2}_{r} \bfX_{.,1:r-1}' \bfX_{1:r-1} + \bfT_{1:r-1,1:r-1}^{-1} )^{-1})$.


\subsection*{Full conditional distribution of common factors vector $\bfx_i$}

Conditional on the $i$th observation belonging to the $k$th cluster, that is $z_i = k$, the full conditional density of $\bfx_i$ is
\begin{align*}
    p(\bfx_i | z_i = k, -) &\propto p(\bfx_i) p(\bfx_i | \bfmu_k, \bfOmega_k , z_i = k)\\
    &\propto exp \left( -\frac{1}{2} \left( \bfx_{i}' \bfB' \bfV^{-1} \bfB \bfx_{i} -2\bfx'_{i} \bfB' \bfV^{-1} \bfy_{i} + \bfx'_{i} \bfOmega_k^{-1} \bfx_{i} -2\bfx'_{i} \bfOmega_k^{-1} \bfmu_k  \right) \right)\\
    &\propto exp \left( -\frac{1}{2} (\bfx'_{i} ( \bfB' \bfV^{-1} \bfB + \bfOmega^{-1}_k ) \bfx_{i} -2\bfx'_{i} (\bfB' \bfV^{-1} \bfy_{i} + \bfOmega_k^{-1} \bfmu_k) ) \right).
\end{align*}

Therefore, the full conditional distribution of $\bfx_i$ when $z_i = k$ is  $N((\bfOmega_k^{-1} + \bfB' \bfV^{-1} \bfB)^{-1} (\bfB'\bfV^{-1}\bfy_{i} + \bfOmega_k^{-1} \bfmu_k), (\bfOmega_k^{-1} + \bfB' \bfV^{-1} \bfB)^{-1})$, $i=1,\ldots,n$.

\subsection*{Full conditional distribution of cluster mean vector $\bfmu_k$}

Recall that $C_k$ is the set of observations in cluster $k$ and that $n_k$ is the number of observations in $C_k$. Then, the full conditional density of $\bfmu_k$ is
\begin{align*}
    p(\bfmu_k | -) &\propto p(\bfmu_k) \prod_{i \in C_k} p(\bfx_i | \bfmu_k, \bfOmega_k, z_i = k)\\
    &\propto exp \left( -\frac{1}{2} \sum_{i \in C_k} (\bfx_{i} - \bfmu_k)' \bfOmega_k^{-1} (\bfx_{i} - \bfmu_k)\right) exp \left( -\frac{1}{2} \left( \bfmu_k - \bfm_k \right)' \bfC_k^{-1} \left( \bfmu_k - \bfm_k \right) \right)\\
    &\propto exp \left( -\frac{1}{2} \bfmu_k' \left( n_k \bfOmega^{-1}_k + \bfC_k^{-1} \right) \bfmu_k -\bfmu_k \left( \bfOmega_k^{-1} \sum_{i\in C_k} \bfx_{i} + \bfC_k^{-1} \bfm_k \right) \right).
\end{align*}

Therefore, the full conditional of $\bfmu_k$ is $N ( ( \bfC_k^{-1} + n_k \bfOmega_k^{-1} )^{-1} \left( \bfC_k^{-1} \bfm_k + n_k \bfOmega_k^{-1} \bar{\bfx}^*_k \right), \left( \bfC_k^{-1} + n_k \bfOmega_k^{-1} \right)^{-1} )$, where $\bar{\bfx}^*_k = \sum_{i \in C_k} \bfx_i / n_k$.

\subsection*{Full conditional distribution of the $k$th cluster covariance $\bfOmega_k$}

Recall that the covariance matrix of the first cluster $\bfOmega_1$ is assumed to be diagonal. 
Consider the $l$th diagonal element $\omega_{1l}$ of $\bfOmega_1$, $l = 1, \dots, F$. The full conditional density of $\omega_{1l}$ is
\begin{align*}
    p(\omega_{1l} | -) &\propto p(\omega_{1l}) \prod_{i \in C_1} p(\bfx_i | \bfmu_1, \omega_{1l}, z_i = 1)\\
    &\propto \omega_{1l}^{-n_1/2 - n_\omega/2 - 1} exp \left( -\frac{1}{2 \omega_{1l}} \left( \sum_{i \in C_1} (x_{il} - \mu_{1l})^{2} + n_\omega s^2_\omega \right) \right).
\end{align*}

Thus, the full conditional distribution of $\omega_{1l}$ is $IG((n_1 + n_\omega)/2 , (\sum_{i \in C_1} (x_{il} - \mu_{1l})^{2} + n_\omega s^2_\omega)/2)$.

For $k > 1$, the full conditional density of $\bfOmega_k$ is 
\begin{eqnarray*}
    p(\bfOmega_k | -) &\propto & p(\bfOmega_k) \prod_{i \in C_k} p(\bfx_i | \bfmu_k, \bfOmega_k, z_i = k)\\
    &\propto & |\bfOmega_k|^{-(n_k + \nu + F + 1)/2} exp \left( -\frac{1}{2} \left( tr \left( \bfOmega_k^{-1} \sum_{i \in C_k} (\bfx_{i} - \bfmu_k) (\bfx_{i} - \bfmu_k)' \right) + tr(\bfOmega_k^{-1} \bfPsi_k) \right) \right)\\
    &\propto & |\bfOmega_k|^{-(n_k + \nu + F + 1)/2} exp \left( -\frac{1}{2} \left( tr \left( \bfOmega_k^{-1} \left( \sum_{i \in C_k} (\bfx_{i} - \bfmu_k) (\bfx_{i} - \bfmu_k)' + \bfPsi_k \right) \right) \right) \right).
\end{eqnarray*}

Thus, the full conditional distribution of $\bfOmega_k$ is $IW(n_k + \nu, \sum_{i \in C_k} (\bfx_{i} - \bfmu_k) (\bfx_{i} - \bfmu_k)' + \bfPsi_k)$.

\subsection*{Full conditional distribution of factor loadings variance $\tau_l$}

Let $\bfb_{.l}^*$ be the $l$th column of $\bfB$ without the first $l$ elements. Then, the full conditional distribution of $\tau_l$ can be obtained by
\begin{eqnarray*}
p(\tau_l | - ) &\propto & p(\tau_l) p(\bfb_{.l}^* | \tau_l)\\
& \propto & \frac{n_\tau s^2_\tau/2}{\Gamma(n_\tau/2)} \tau_l^{-(n_\tau/2 + 1)} exp \left( -\frac{n_\tau s_\tau^2 / 2}{\tau_l}\right) \tau_l^{-(R - l)/2}exp \left( -\frac{1}{2 \tau_l} \bfb^{*'}_{.l} \bfb^{*}_{.l}  \right)\\
& \propto & \tau_l^{-(R - l + n_\tau)/2  - 1} exp \left( -\frac{1}{2 \tau_l} (\bfb_{.l}^{*'} \bfb_{.l}^{*} + n_\tau s_\tau^2) \right).
\end{eqnarray*}
Therefore, the full conditional of $\tau_l$ is $IG((R -l + n_\tau)/2, (\bfb_{.l}^{*'} \bfb_{.l}^{*} + n_\tau s^2_\tau)/2)$.

\subsection*{Full conditional distribution of idiosyncratic variance $\sigma_r^2$}

The full conditional of the $r$th idiosyncratic variance $\sigma^2_r$ can be obtained by
\begin{eqnarray*}
    p(\sigma^2_r | -) &\propto & p (\sigma^2_r) f(\bfy_{.r} | \bfX, \bfb_r, \sigma^2_r) \\
    &\propto & (\sigma^2_r)^{-n/2} exp \left( -\frac{1}{2 \sigma^2_r} \sum_{i=1}^n (y_{ir} - \bfx_i' \bfb_r)^2 \right)\\
    &\propto & (\sigma^2_r)^{-(n + n_\sigma)/2 - 1} exp \left( -\frac{1}{2 \sigma^2_r} \left( \sum_{i=1}^n (y_{ir} - \bfx_i' \bfb_r)^2 + n_\sigma s^2_\sigma \right) \right).
\end{eqnarray*}

Therefore, the full conditional of $\sigma^2_r$ is $IG ((n + n_\sigma)/2,( \sum_{i=1}^n (y_{ir} - \bfx_i' \bfb_r)^2 + n_\sigma s^2_\sigma)/2))$.


\subsection*{Full conditional distribution of the vector of cluster probabilities $\bfp = (p_1, \dots, p_K)$}

The full conditional density of the vector of cluster probabilities $\bfp = (p_1, \dots, p_K)$ is
\begin{eqnarray*}
    p(\bfp | -) &\propto & p(\bfp) p(\bfZ|\bfp)
    \ \propto \ \prod_{k = 1}^K p_k^{n_k} \prod_{k = 1}^K p_k^{\alpha_k}
    \ \propto \ \prod_{k = 1}^K p_k^{n_k + \alpha_k}.
\end{eqnarray*}

Therefore, the full conditional distribution of $\bfp$ is $Dirichlet(n_1 + \alpha_1, \dots, n_K + \alpha_K)$.

\end{document}


\maketitle
\begin{figure}
\centering
\includegraphics[width=8cm]{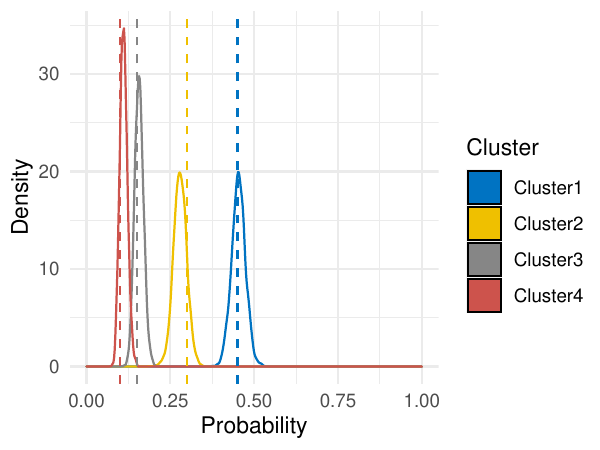}
\caption{Simulated data -- posterior densities (solid lines) of the cluster probabilities for BCFM with $K = 4$ clusters and $F = 3$ factors. For comparison, vertical dashed lines indicate the true values of the cluster probabilities.}\label{SD_Probability_Density}
\end{figure}
\newpage

\begin{figure}
\centering
\includegraphics[width=15cm]{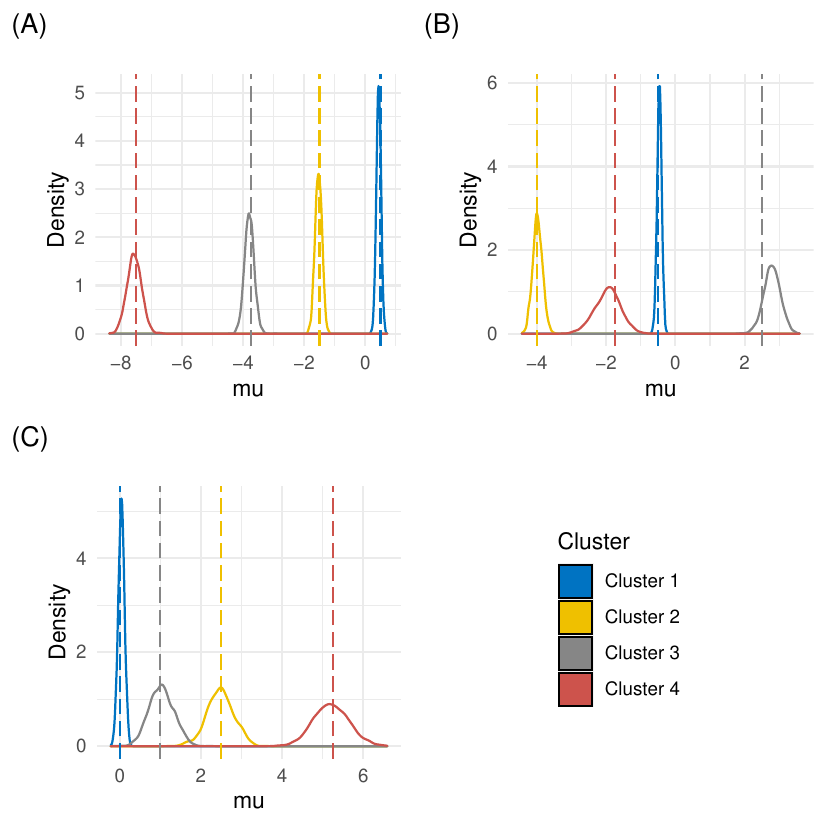}
\caption{Simulated data -- posterior densities of the elements of the mean vectors for the common factors of each cluster. (A--C) Each panel corresponds to the means of a common factor across clusters. (A) first factor, (B) second factor, and (C) third factor. For comparison, vertical dashed lines indicate the true values of the means of the common factors within each cluster.}\label{SD_mu_Density}
\end{figure}
\newpage

\begin{figure}
\centering
\includegraphics[width=15cm]{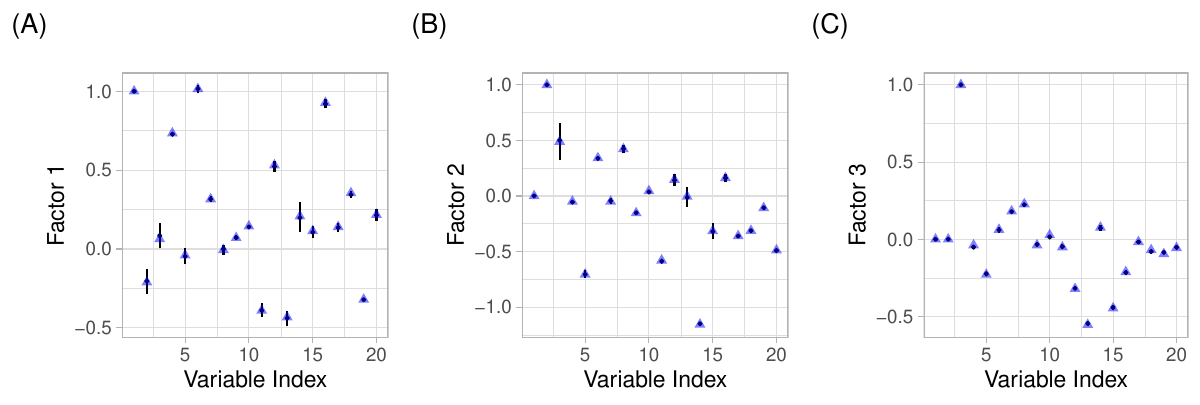}
\caption{Simulated data -- posterior summaries of factor loadings for BCFM with $K = 4$ clusters and $F = 3$ factors: true value (blue triangle), posterior mean (black circle), and 95\% credible interval (black vertical line). (A) first factor, (B) second factor, and (C) third factor.}\label{SD_B_Density}
\end{figure}

\newpage

\begin{figure}
\centering
\includegraphics[height=8cm]{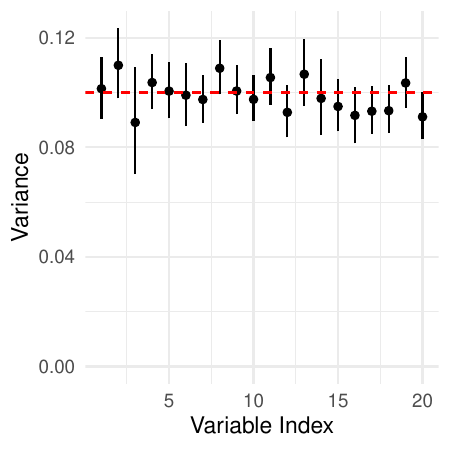}
\caption{Simulated data -- idiosyncratic variances for BCFM with $K = 4$ clusters and $F = 3$ factors: 95\% credible interval (vertical line) and posterior mean (circle), and true values (red dashed line).}\label{SD_sigma2_Density}
\end{figure}

\newpage

\begin{figure}
\centering
\includegraphics[width=10cm]{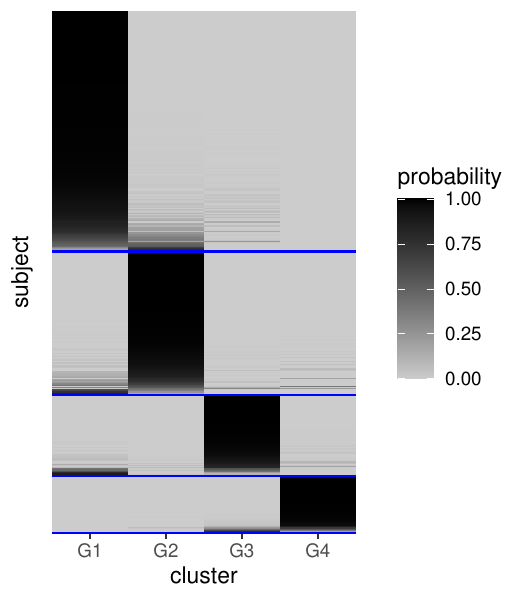}
\caption{Simulated data -- heatmap of the cluster assignment probabilities and the true clusters for BCFM with $K = 4$ clusters and $F = 3$ factors. Blue lines present the boundaries of the true clusters. The shades represent the posterior probability that each subject belongs to each cluster.}\label{SD_heatmap}
\end{figure}

\newpage

\begin{figure}
\centering
\includegraphics[width=8cm]{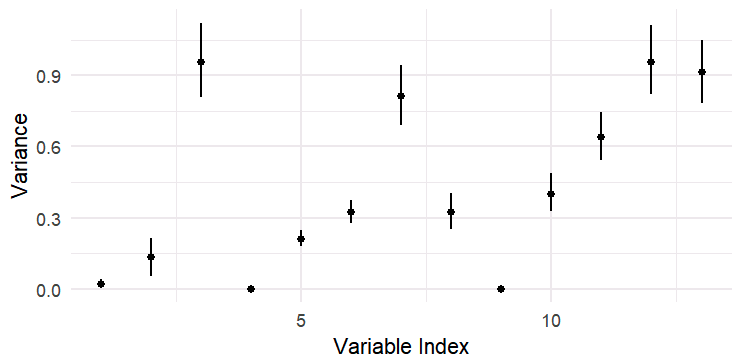}
\caption{OUD Recovery data -- idiosyncratic variance for the 5-cluster-4-factor BCFM: posterior mean (black circle), and 95\% credible intervals (black line).}\label{BL_sigma2_Density}
\end{figure}